\documentclass[aps,prl,showpacs]{revtex4}
\usepackage{epsfig}
\usepackage[dvips]{color}     
\usepackage[latin1]{inputenc} 
\usepackage{graphicx}
\usepackage{amssymb}
\usepackage{caption}
\DeclareCaptionFormat{}{#3}
\newcommand{\linea}{\noindent\mbox{}\hrulefill \\}

\newcommand{\be}{\begin{equation}}
\newcommand{\ee}{\end{equation}}
\newcommand{\bea}{\begin{eqnarray}}
\newcommand{\eea}{\end{eqnarray}}
\newcommand{\bt}{\begin{tabular}}
\newcommand{\et}{\end{tabular}}
\newcommand{\ba}{\begin{array}}
\newcommand{\ea}{\end{array}}

\newcommand{\pr}{{\rm I}}
\newcommand{\se}{{\rm II}}
\newcommand{\um}{{1\over 2}}

\begin{document}

\title{\mbox{Second discontinuity in the specific heat of two-phase superconductors}}

\author{E. Di Grezia}
\email{digrezia@na.infn.it}
\affiliation{\mbox{Universit\`a Statale di Bergamo, Facolt\`a di
Ingegneria}\\
viale Marconi 5, 24044 Dalmine (BG), Italy\\
\mbox{and Istituto Nazionale di Fisica Nucleare, Sezione di Milano}
\\
via G. Celoria 16, I-20133 Milan, Italy}

\author{S. Esposito}
\email{salvatore.esposito@na.infn.it}
\affiliation{\mbox{Dipartimento di Scienze Fisiche, Universit\`{a} di
Napoli ``Federico II''}
\mbox{and Istituto Nazionale di Fisica Nucleare, Sezione di Napoli,}\\
Complesso Universitario di Monte S. Angelo, Via Cinthia, I-80126 Napoli, Italy}

\author{G. Salesi}
\email{salesi@unibg.it}
\affiliation{\mbox{Universit\`a Statale di Bergamo, Facolt\`a di
Ingegneria}\\
viale Marconi 5, 24044 Dalmine (BG), Italy\\
\mbox{and Istituto Nazionale di Fisica Nucleare, Sezione di Milano}
\\
via G. Celoria 16, I-20133 Milan, Italy}



\begin{abstract}
The recently proposed theoretical model of superconductors endowed with two distinct superconducting phases and critical temperatures is further analyzed by introducing two distinct order parameters, described by two scalar fields which condensate at different temperatures. We find some deviations in basic thermodynamical quantities with respect to the Ginzburg-Landau one-phase superconductors. In particular, in contrast to the usual case where  only one jump in specific heat takes place at the normal-superconductor transition temperature, we actually predict an additional discontinuity for $C_V$ when passing from a superconducting phase to the other one.


\end{abstract}

\maketitle

In the Ginzburg-Landau (GL) theory\cite{GL,Annett} the superconductivity
is described in terms of a complex order parameter $\phi$, which can be interpreted as the
wave function of the Cooper pair in its center-of-mass frame. 
The classical phenomenological GL approach entails a (unique) critical temperature $T_C$, without assuming a particular temperature-dependence 
of the coefficient $a(T)$ appearing in the effective free energy function for unit volume, expanded up to the $|\phi|^4$ order:
\be
F \simeq F_{\rm n} + a(T)|\phi|^2 + \frac{\lambda}{4}\,|\phi|^4\,. 
\label{Effe}
\ee
Quantity $F_{\rm n}$ indicates the normal-phase (not superconducting) free energy density; \  while $\lambda$, giving the strength of the Cooper pair binding, is assumed to be approximately constant. Ginzburg and Landau only assumed that the coefficient $a(T)$ is positive above $T_C$, vanishes when the temperature approaches the critical value and becomes negative for $T<T_C$; around the critical temperature changes very smoothly: 
\ $a(T) \simeq \dot{a}(T_C)(T-T_C)$\,. In the alternative quantum field approach
analyzed below we instead adopt well-defined analytic expressions for
$a(T)$ as a function of the temperature.

Actually, because of the interaction of the charged scalar field $\phi$ with the  electromagnetic field $A^\mu$, the order parameter is usually associated to the Higgs field responsible of the U(1) spontaneous symmetry breaking (SSB) \cite{Higgs,Bailin,Tinkham} occurring during the 
normal state-superconducting phase transition.
As a consequence of the symmetry breaking, due to a non-vanishing expectation value of the order parameter in the ground state below the critical temperature, the photon acquires a mass (causing the Meissner effect) and the system becomes superconducting. 
By adopting this approach, we can 
initially start from a relativistically invariant Lagrangian containing the interaction of a single $\phi$ with $A^\mu$ as well as the $\lambda$ self-interaction (hereafter $\hbar=c=1$):
\begin{equation} 
{\cal L}=\left( D_{\mu }\phi \right)^{\dagger }\left( D^{\mu }\phi
\right) + m^2\phi^\dagger\phi - \frac{\lambda}{4}(\phi^\dagger\phi)^2
-\frac{1}{4}F_{\mu\nu}F^{\mu\nu}\,, \label{Elle}
\end{equation}
where \ $m^2>0$, 
\ $F_{\mu\nu}\equiv\partial_\mu A_\nu - \partial_\nu A_\mu$
is the electromagnetic field strength,
\ and \ $D_{\mu }\equiv\partial_{\mu }+2ieA_{\mu }$ \ is the covariant derivative 
($2e$ is the electric charge of a Cooper pair).
Given the above Lagrangian, the effective free energy density at finite temperature 
is formally identical to the GL expression given in (\ref{Effe}). 
However, despite the outstanding importance of the GL theory in superconductivity, as well as in other physical systems, it has still not been solved exactly beyond the mean-field approximation. Whilst this was not a serious problem for traditional superconductors, where the Ginzburg temperature interval is small around the critical temperature, the situation has changed especially with the advent of high-$T_c$ superconductors. In fact, for these systems, the Ginzburg temperature interval is large and
we may expect strong field fluctuations and critical properties beyond the mean-field approximation. Indeed, in high-$T_c$ superconductors several experiments have observed critical effects in the specific heat \cite{critical}, although the presence of a magnetic field generally makes the situation more complicated. On a theoretical side, the effect of gauge field fluctuations causes great difficulties in the critical phenomena theory and, unlike the simpler $\phi^4$ theory for a neutral superfluid, the exact critical behaviour remains unknown. It is well-known that at the mean-field level the superconductive transition is discontinuous, but it seems that this result is confirmed even when field fluctuations are included \cite{fluct}. This is also confirmed by numerical simulations of lattice models for small values of the Ginzburg-Landau parameter $\varkappa$, while for large $\varkappa$ the results are consistent with a continuous, second-order phase transition \cite{second}. It is thus a general belief that the standard GL model leads to a first-order transition instead of a continuous transition, but several other studies at one-loop (and even at two-loop) approximation have been carried out in recent years (see, for example, \cite{one1, one2, one3, one4} and references therein). Some of them entail runaway solutions of the GL equations (pointing towards first-order transitions), while others find a scaling behaviour with a new stable fixed point in the space of static parameters. Also, again beyond the plain mean-field approximation, Kleinert has shown the existence of a tricritical point in a superconductor, by taking the vortex fluctuations into account.
In a recent paper \cite{TwoTC} we also considered one-loop radiative corrections to the classical GL theory and obtained, according to the chosen ``condensation gauge'', {\em two distinct} well-defined expressions of the coefficient $a$ as a function of $T$, and, correspondingly, {\em two different critical temperatures} $T_1$ and $T_2$.
Let us start by expanding a complex field $\phi$ as follows
\begin{equation}
\phi \equiv \frac{1}{\sqrt
2}(\eta_0+\eta)\,{\rm e}^{i\theta/\eta_0}\,, 
\label{Phi_I}
\end{equation}
where $\eta_0$ is a real constant, $\eta$ and $\theta$ are real fields.
Then, if we let the scalar field fluctuate around the minimum of the free energy, a condensation of the field $\eta$ takes place as a result of the $U(1)$ SSB. In Eq.\,(\ref{Phi_I}) the constant field $\eta_0/\sqrt{2}$ is defined as the expectation value (the condensation value) of the modulus of the scalar field $\phi$. \ Finite-temperature one-loop quantum corrections to the $T=0$ expression of the free energy density lead to \cite{NXC}
\be
F_\pr = F_{\rm n} + \um a_\pr(T)\eta_0^2 + \frac{\lambda}{16}\,\eta_0^4 
\ee
with
\be
a_\pr = - m^2 + \frac{\lambda + 4e^2}{16}\,T^2\,.
\label{a_I}
\ee
The parameter $a_\pr$ vanishes when the temperature approaches a critical value
given by  
\be
T_1 = 2\sqrt{\frac{4m^2}{\lambda + 4e^2}}\,. 
\label{T1}
\ee
Below $T_1$ the expectation value of $\eta_0^2$ which minimizes the free
energy function results to be
\be
\eta_0^2(T) = -\frac{4a_\pr(T)}{\lambda}\,.
\label{etaquad}
\ee
Alternatively, we may expand the field $\phi$ as:
\begin{equation}
\phi \equiv \frac{1}{\sqrt 2}(\phi_0 + \phi_{a} +i \phi_{b}),
\label{Phi_II}
\end{equation}
where $\phi_0$ is a real constant, and $\phi_{a}, \phi_{b}$ are two real scalar fields. Now we assume that a condensation takes place in the field $\phi_a$ (or, equivalently, in $\phi_b$) rather than in the component $\eta$. In Eq.\,(\ref{Phi_II}) the constant field $\phi_0/\sqrt{2}$ is defined as the  expectation value of the real part of $\phi$. In this case, after such condensation, the effective Helmholtz energy density writes 
\be
F_\se = F_{\rm n} + \um a_\se(T)\phi_0^2 + \frac{\lambda}{16}\,\phi_0^4 
\ee
with \cite{Bailin}
\be
a_\se = - m^2 + \frac{\lambda + 3e^2}{12}\,T^2\,.
\label{a_II}
\ee
From the vanishing of $a_\se$ we now derive a \emph{different} critical temperature 
\be
T_2 = 2\sqrt{\frac{3m^2}{\lambda + 3e^2}}\,.
\label{T2}
\ee
Since \ $\infty>\lambda>0$, \ we correspondingly have \ $\displaystyle \frac{\sqrt{3}}{2}\,T_1<T_2<T_1$. \
Accordingly, for very large self-interaction, \ $\lambda/e^2\to\infty$, \
\emph{we predict a maximum difference of} 15\% \emph{between the two critical temperatures} \cite{TwoTC}.

Below $T_2$ the expectation value for $\phi_0^2$ which minimizes the free
energy function is given by
\be
\phi_0^2(T) = -\frac{4a_\se(T)}{\lambda}\,.
\label{phiquad}
\ee
We understand the appearing of a new lower critical temperature
when expanding the exponential in Eq.\,(\ref{Phi_I}) in $\theta/\eta_0$ and comparing with Eq.\,(\ref{Phi_II}):
\begin{eqnarray}
\phi_0 & \sim & \eta_0
\nonumber \\
\phi_a & \sim & \eta - \frac{\theta}{2}\left(\frac{\theta}{\eta_0}\right)+ \cdots\label{A5}\\
\phi_b & \sim & \theta \, + \, \eta\left(\frac{\theta}{\eta_0}\right)+ \cdots\,.
\nonumber
\end{eqnarray}
The degrees of freedom carried out by the real scalar fields $\phi_a ,\phi_b$ are different from those corresponding to $\eta, \theta$, and tend to coincide only in the limit $\eta_0\rightarrow \infty$. Actually, in Eqs.\,(\ref{A5}) the higher orders in $\eta_0^{-1}$ contribute at the denominator of the expression (\ref{T1}) as an additional $\lambda/3$ term; that is an
increased effective self-interaction of the Cooper pairs arises ($\lambda\rightarrow
\lambda_{\rm eff}=4\lambda/3$) \cite{TwoTC}.

Since, as we have seen, two different condensations are allowed to occur inside the same system, we do not a priori exclude any of them. Hence we are led to introduce two order parameters, that is two scalar charged fields: the first one related to the condensation of the modulus of $\phi_\pr$ (the corresponding phase will be hereafter denominated as ``phase I''); while the second one related to the condensation of the real part of $\phi_\se$ (``phase II'').

Neglecting possible interactions between the two scalar fields, the total Lagrangian now writes:
	\be
	{\cal L} =\left(D_{\mu }\phi_\pr \right)^{\dagger }\left( D^{\mu }\phi_\pr
	\right) + m^2\phi_\pr^\dagger\phi_\pr - \frac{\lambda}{4}(\phi_\pr^\dagger\phi_\pr)^2+
	\left(D_{\mu }\phi_\se \right)^{\dagger }\left( D^{\mu }\phi_\se
	\right) + m^2\phi_\se^\dagger\phi_\se - \frac{\lambda}{4}(\phi_\se^\dagger\phi_\se)^2 -
	\frac{1}{4}F_{\mu\nu}F^{\mu\nu}\,.
	\label{Ellex2}
	\ee
As a matter of fact, starting from high values and then lowering the temperature we meet a first SSB at the critical temperature $T_1$: the medium becomes superconducting. 
Since the II-phase term $a_\se(T){\phi_0}^2 + \lambda\,{\phi_0}^4$ in the free energy density is negative for $T<T_2$, by further lowering the temperature at $T=T_2$ the condensation involving the second order-parameter is energetically favored and a new (second-order) phase transition starts.
Below $T_2$ the system is ``more'' superconducting with respect to the GL standard case since, in addition to the phase-I Cooper pairs, we should observe also the formation of phase-II Cooper pairs.
Such two superconducting phases correspond to different condensations of electrons in Cooper pairs which exhibit different self-interaction, and are described by different scalar fields. The realization of one of the two regimes is ruled by the relative strength of the Cooper pair
self-interaction ($\lambda$) with respect to the electromagnetic interaction ($e$).

Correspondingly, the total free energy density, being an additive quantity, results
as the sum of contributions from normal-conducting electrons, phase-I superconducting Cooper pairs, and phase-II superconducting Cooper pairs:
\begin{eqnarray}
F = F_{\rm n} & \ \mbox{\rm for} \ & T>T_1\,, 
\label{Norm}\\
F = F_{\rm n} + \um a_\pr(T){\eta_0}^2 + \frac{\lambda}{16}\,{\eta_0}^4 &
\ \mbox{\rm for} \ & T_2<T<T_1\,, \ \ \
\label{I}\\
F = F_{\rm n} + \um a_\pr(T){\eta_0}^2 + \frac{\lambda}{16}\,{\eta_0}^4 +
\um a_\se(T){\phi_0}^2 + \frac{\lambda}{16}\,{\phi_0}^4 & \ \mbox{\rm for} \ & T<T_2\,,
\label{I+II}
\end{eqnarray}
($\eta_0$ indicates the expectation value of $|\phi_\pr|$; $\phi_0$ indicates the expectation value of ${\rm Re}\{\phi_\se\}$).

From Eqs.(\ref{T1}) and (\ref{T2}) we are able to put the two free parameters of our theory, i.e. the ``mass squared'' $m^2$ and the self-interaction coupling constant $\lambda$ as functions of the two critical temperatures:
\be
m^2 = \frac{e^2T_1^2T_2^2}{4(4T_2^2-3T_1^2)}\,,
\ee
\be
\lambda = \frac{12e^2(T_1^2-T_2^2)}{4T_2^2-3T_1^2}\,.
\ee
Therefore experimental measurements of $T_1$ and $T_2$ could yield an estimate of the dynamical parameters ruling the SSB and the electron binding in Cooper pairs. Notice that such a goal is not possible in the framework of the standard GL, theory where the parameters in (\ref{Effe}) are not explicitly determined.

Inserting the above expressions in (\ref{a_I}), (\ref{etaquad}), (\ref{a_II}), and (\ref{phiquad}), also the expectation values of the two scalar fields can be expressed in terms of $T_1$ and $T_2$:
\be
\eta_0^2(T) = \frac{T_2^2(T_1^2-T^2)}{12(T_1^2-T_2^2)}\,,
\ee
\be
\phi_0^2(T) = \frac{T_1^2(T_2^2-T^2)}{12(T_1^2-T_2^2)}\,.
\ee
By inserting in (\ref{I}) and (\ref{I+II}) we may compare, for $T<T_2$, the behavior of the free energy in the GL case, where (\ref{I}) holds also for $T<T_2$, and in the case of two-phases superconductors for which, instead, (\ref{I+II}) applies. The free energy difference results to be
\be
\Delta F \equiv F_{\rm GL} - F_{\rm 2ph} 
= \frac{e^2T_1^4(T_2^2-T^2)^2}{48(4T_2^2-3T_1^2)(T_1^2-T_2^2)}\,.
\label{Delta F}
\ee
We see that such a difference increases by lowering the temperature and reaches its 
maximum for $T=0$.

The pressure is given by \ $\displaystyle P=-\left.\frac{\partial{\cal F}}{\partial V}\right|_T$, \ where ${\cal F}=FV$ is the free energy. Since that the superconductive part of the free energy density is independent of the volume, we have
\be
\Delta P \equiv P_{\rm GL} - P_{\rm 2ph} = - \Delta F < 0\,.
\ee
Hence the pressure is expected to be larger for two-phases superconductors.
Thus the differences in the free energy and in the pressure become
more sensible far from $T_2$, near to absolute zero. 

From \ $\displaystyle S = -\left.\frac{\partial F}{\partial T}\right|_{V}$, \ for $T<T_2$, we get the difference in the entropy density:
\be
\Delta S \equiv S_{\rm GL} - S_{\rm 2ph} 
= \frac{e^2T_1^4T(T_2^2-T^2)}{12(4T_2^2-3T_1^2)(T_1^2-T_2^2)}\,.
\ee
Being $\Delta S>0$, we can say that the two-phases superconductors are in a sense more ``ordered'' than the GL ones, the maximum difference for the entropy being reached
at \ $T=T_2/\sqrt{3}$\,.

\begin{figure}
\centering
\includegraphics[scale=0.7]{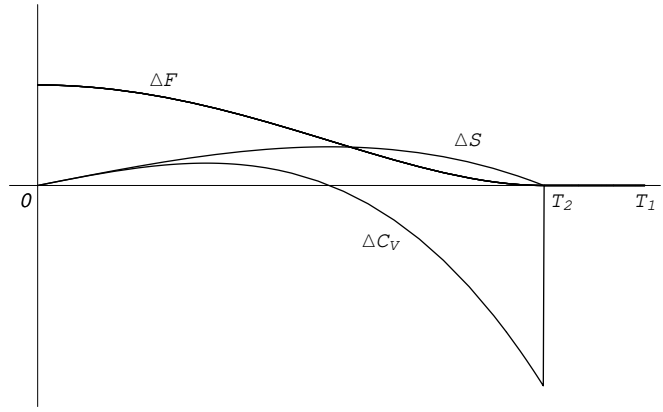}   
\caption{Differences between GL and two-phases superconductors}
\label{}
\linea
\end{figure}

We can compare as well the latent heat absorbed during the formation of the superconducting phase in GL and two-phase superconductors at a given temperature $T<T_2$ ($S_0$ indicates the 
entropy of the normal-phase)
$$
\lambda_{\rm GL}(T) = T(S_0 - S_{\rm GL}) =
$$
\be 
= \frac{e^2T_2^4T^2(T_1^2-T^2)}{12(4T_2^2-3T_1^2)(T_1^2-T_2^2)}\,\,;
\ee
$$
\lambda_{\rm 2ph}(T) = T(S_0 - S_{\rm 2ph}) = 
$$
\be
= \frac{e^2T^2[T_2^4(T_1^2-T^2) + T_1^4(T_2^2-T^2)]}{12(4T_2^2-3T_1^2)(T_1^2-T_2^2)}\,.
\ee
The difference between $\lambda_{\rm GL}$ and $\lambda_{\rm 2ph}$ reaches its maximum at $T=T_2/2$.

Finally, applying the well-known formula for the specific heat at constant volume
\be
C_V = T\,\left.\frac{\partial S}{\partial T}\right|_V\,,
\ee
we obtain the difference in $C_V$ ($T<T_2$)
\be
\Delta C_V \equiv C_{V_{\rm GL}} - C_{V_{\rm 2ph}}
=\frac{e^2T_1^4T(T_2^2-3T^2)}{12(4T_2^2-3T_1^2)(T_1^2-T_2^2)}\,,
\ee
which is positive for \ $0<T<T_2/\sqrt{3}$, \ negative for \ $T_2/\sqrt{3}<T<T_2$, \ and vanishes at \ $T=T_2/\sqrt{3}$. \
As it is seen in the figure, whilst $\Delta F(T_2)$ and $\Delta S(T_2)$ vanish, so that $F$ and $S$ are continuous in $T_2$, quantity $\Delta C_V(T_2)$ is not zero: 
\be
\Delta C_V(T_2) = -\,\frac{e^2T_1^4T_2^3}{6(4T_2^2-3T_1^2)(T_1^2-T_2^2)}<0\,.
\ee
We then observe a finite jump in the specific heat also in the transition from the  superconducting first phase (I) to the second one (II), while in GL superconducting media only
one discontinuity is expected (for $T=T_1$, when the system changes from the normal to the superconducting regime). This sudden change in the heat capacity is a distinguishing characteristic of a first order phase transition. Since the jump of the specific heat in $T=T_1$ results to be
\be
\Delta C_V(T_1) =-\,\frac{e^2T_2^4T_1^3}{6(4T_2^2-3T_1^2)(T_1^2-T_2^2)}\,,
\ee
the ratio between the two discontinuities can be written as
\be
\frac{\Delta C_V(T_2)}{\Delta C_V(T_1) } = \frac{T_1}{T_2}\,.
\ee
Being the above ratio larger than 1 (and smaller than $\sqrt{4/3}$\,\,), we expect the two
jumps to be comparable. Thus also the second jump at the lower temperature ---just a novel effect because it happens between two superconducting phases--- could be experimentally investigated and measured. Notice also that, as expected, both discontinuities increase indefinitely in the large Cooper pairs self-interaction limit, \ $\lambda/e^2\to \infty$, \ $(T_1/T_2)^2\to 4/3$.

A thermodynamical behavior qualitatively similar to the one predicted by our model has been recently observed in MgB$_2$, as it is readily realized by a comparison with the existing experimental literature (see, for example \cite{BuzeaYamashita} and Refs.\,therein). Several two-band theories \cite{2Gaps} try to explain the second discontinuity in the specific heat of MgB$_2$: probably it exists a correspondence between the \textsl{two} ``classical-macroscopic'' (since represent ``collective'' wave-functions for the condensate) order parameters in GL-like approaches as the present one, and the \textsl{two}  ``quantum-microscopic'' gaps in quasi-particle energy spectra predicted for MgB$_2$ by some BCS-like theories.

\

We stress that our results, as already pointed out, are a direct consequence of one-loop field theory calculations of the GL parameters. While computations beyond the plain mean-field approximation and their physical consequences have been largely considered and debated in the literature \cite{one1, one2, one3, one4}, nevertheless those authors did not consider further degrees of freedom owned by a mean-field theory as the present one, which entails two concurrent order parameters and two differently bonded Cooper pairs. Consequently in the above-mentioned literature we do not find predictions of a second discontinuity in the specific heat or of some anomalous magnetic effects \cite{Magn2TC}.  However, it is highly desirable that even more attention will be paid in the near future to explore the striking physical phenomena coming out from GL-like theories when field fluctuations are fully taken into account.

\

Finally, let us remark that attractive interactions (``Cooper-effect") and ``gapped" energy spectrum 
characterize both quantum theory of {\em fermionic} superfluids (as e.g. $^3$He) and BCS theory of superconductivity. Consequently we might expect that the basic thermodynamical properties here found for two-phase superconductors could analogously apply to a sort of ``two-phase Fermi superfluids'' (endowed with two critical temperatures) as well.

\

\

\end{document}